\documentclass{mem}
\usepackage{natbib}\usepackage{txfonts}\usepackage{balance}
\usepackage{flushend}
\usepackage{graphicx}
\usepackage{placeins}
\usepackage[a4paper,breaklinks,pdftex]{hyperref}
\idline{75}{282}
\begin{document}
\def\teff{$T\rm_{eff }$}
\def\kms{$\mathrm {km s}^{-1}$}

\title{
Exploring the Morphologies of High Redshift Galaxies with Machine Learning
}

   \subtitle{}

\author{
Clár-Bríd\,Tohill\inst{1,3} 
\and Steven\, Bamford\inst{1}
\and Christopher\, Conselice\inst{2}
          }

\institute{
University of Nottingham,
 School of Physics \& Astronomy, 
 Nottingham, NG7 2RD, UK
\and
Jodrell Bank Centre for Astrophysics, 
University of Manchester, Oxford Road, Manchester UK
\and
Isaac Newton Group of Telescopes, 
38700 Santa Cruz de La Palma, Canary Islands, Spain
\email{cbtohill@ing.iac.es}
}

\authorrunning{Tohill }

\titlerunning{Galaxy morphology with ML}

\date{Received: Day Month Year; Accepted: Day Month Year}

\abstract{
The morphology of a galaxy has been shown to encode the evolutionary history and correlates strongly with physical properties such as stellar mass, star formation rates and past merger events. While the majority of galaxies in the local universe can be classified on the Hubble sequence, little is known about the different types of galaxies we observe at high redshift. The irregular morphology of these galaxies makes visual classifications difficult, and with the future of astronomy consisting of many `Big Data' surveys we need an efficient, and unbiased classification system in place. In this work we explore the use of unsupervised machine learning techniques to preform feature extraction from galaxy images to separate high redshift galaxies into different morphological types based on the machine learning clusters. We expand on previous work by addressing observation biases such as the orientation, apparent size of the galaxies and noise before extracting features, thus reducing the number of clusters and forcing the network to learn meaningful features. We then compare the extracted clusters’ physical properties, to investigate the separation between the groups. 
\keywords{Astronomy, Galaxies - 
Machine Learning - Deep Learning - High Redshift }
}
\maketitle{}

\section{Galaxy Morphology}
Galaxy morphology can provide us with a window into understanding its evolution. It has been shown that the morphology encodes the past and ongoing formation modes of a galaxy, which can give us an indication of how galaxies evolved throughout cosmic time \citep{Holmberg, Dressler, Kauffmann}. The Hubble classification scheme describes the morphologies of galaxies observed in the local universe \citep{Hubble}. This classification scheme not only describes the visual appearance of the galaxy, but it has been shown that morphological type correlates strongly with many intrinsic properties such as the SFR, age, number of past merging events etc. 
However, while this system has recently been shown to describe galaxies up to very high redshifts $z \geq 7$(\cite{leo_hubble}), the number of irregular galaxies increases rapidly with redshift, thus requiring an alternative description. \\
\indent Galaxies at high redshift show much more peculiar and clumpy morphologies \citep{Noguchi_Clumpy, conselice_irr}. This is due to many reasons, some we understand such as the cosmic star formation history.We know that the star formation rate in the universe was maximum around $z \sim 2$, meaning that galaxies would have more star forming regions, leading to more clumpy morphologies. 
We also know that the merger rates were higher in these earlier times, thus causing galaxies to show disturbed morphologies with tidal disruptions and multiple cores etc. However what is not understood is how these galaxies evolved to those that we observe today. This evolution can be studied by investigating the evolution of morphology with redshift. 

The problem that we are interested in investigating then is how do we robustly classify these distant galaxies into self-similar types? One solution that has proven successful so far is citizen science projects such as galaxy zoo whereby the general public visually inspects and classifies galaxies according to the presence or absence of certain features \citep{Willett}. While these classifications have been used successfully  in many studies \citep{Schawinski, Bamford, Cardamone}, with future `Big Data' surveys expecting to image billions of galaxies this is unfeasible, and would take many years to amass enough classifications.
There is also an issue of human bias that is imposed when classifying galaxies due to the subjective judgement of the classifier which we do not want when trying to draw conclusions from the results. There is an obvious solution to both these problems and one that has become popular in recent years – machine learning.

\section{Method}

\begin{figure}[t!]
\resizebox{\hsize}{!}{\includegraphics[clip=true]{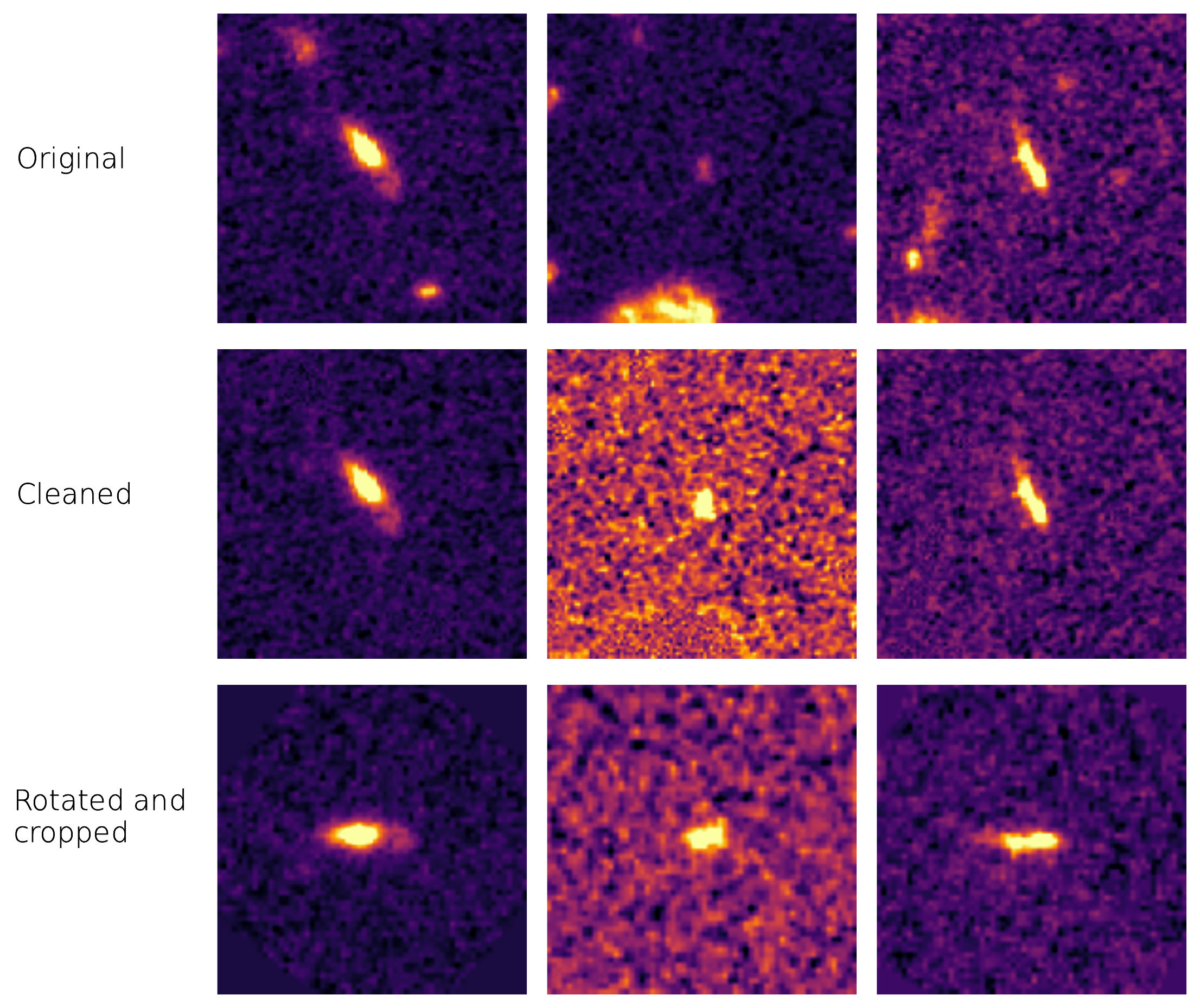}}
\caption{\footnotesize An example of our augmented images. \emph{Top:} Original CANDELS images. \emph{Middle:} This shows the galaxy images after they have been cleaned of background sources. \emph{Bottom:} Re-scaled and rotated images that are used in our network. }
\label{fig:aug}
\end{figure}

In recent years Machine Learning (ML) has proven to be very successful in astronomy and has already been applied to the morphological classification of galaxies with great success. Supervised ML has been used to both identify mergers \citep{Leo_merger} as well as Hubble type classifications \citep{Dieleman, Dominguez_sdss, Sunny}. While these studies have proven to be successful, they require prior knowledge of the data in order to have a labelled training sample. This brings back the issue of human bias within the labels as well as the issue of having enough data to classify. One solution to remove this bias from any future work, and to improve the efficiency of these classifications, is to move towards using unsupervised machine learning (UML) techniques. As the name suggests, UML techniques require no labels to train but uses only the data as an input. For this reason UML methods can be a more robust, and unbiased method for data analysis. 

\begin{figure*}[t!]
\resizebox{\hsize}{!}{\includegraphics[clip=true]{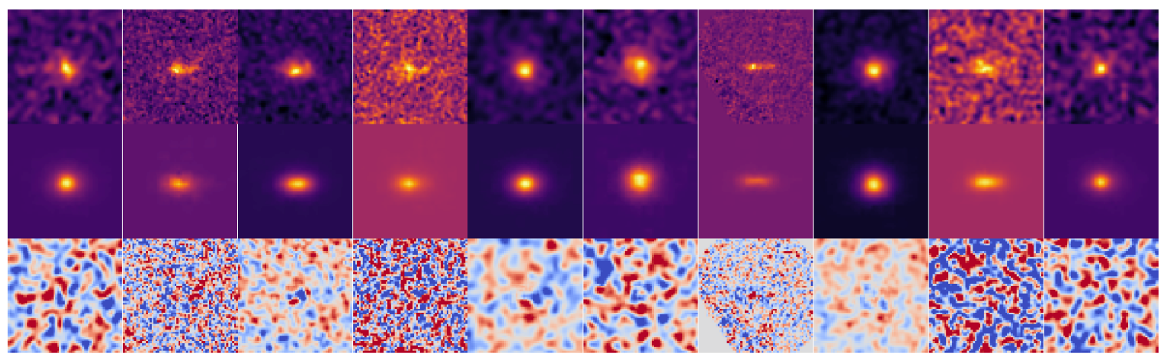}}
\caption{\footnotesize \emph{Top:} Input images to encoder. \emph{Middle:} Reconstructed images using three encoded features. \emph{Bottom:} Residuals showing how the reconstructions have encoded all of the galaxy light and not noise. }
\label{fig:recon}
\end{figure*}

In this work we utilise a type of UML network known as a Variational autoencoder (VAE). The main idea behind an autoencoding network is that of dimensionality reduction. Dimensionality reduction is the process by which the number of features needed to describe some data are reduced. 
A VAE is composed of two main components, the encoder and the decoder. The encoder takes an input, which in this example is an image of a galaxy, and encodes the information into a lower dimensional representation of your data. This lower dimensional representation is stored in the latent space (aka feature space). The decoder then samples from this latent space to create a reconstruction of your input. 
 The VAE is trained to compress your input data whilst minimising the reconstruction loss between it and the output, decoded image. Because you are losing information in this encoding the reconstructions always tend to be smoother, and contain only the main features that describe the data. Data that have similar features will be closer together within this feature space while data with different features will be located in a different region. 
The aim of this work is to encode the main features of our sample of galaxies and investigate the separation between them in this feature space. If we can group points that are close within this latent space we will obtain a selection of galaxies that possess similar morphological features resulting in a broad and robust classification scheme. 

\subsection{Data Augmentation} 
One common issue that can arise with feature extraction is the fact that the network is trying to reproduce the input images with as few features as possible. This causes features such as shape, orientation, size and position to be encoded first as these will result in a smaller reconstruction loss than more finer details. These features however are not intrinsic to the galaxy and are in fact observational biases that we have imposed on the data simply because of our observation position on Earth. 
In our work we want to address and remove these observational biases before trying to cluster our galaxies, thus allowing the feature space to be physically meaningful and without the risk of missing any other subtle features of the galaxies. To do this we need to augment our data.

For our work, we use data from the CANDELS survey as imaged by the Hubble Space Telescope. In total we have 30488 postage stamp galaxy images from $2 \leq z \leq 7$. 
The augmentation steps we apply to our data can be seen in Fig.\ref{fig:aug}. First we remove background sources from our images using the \textsc{galclean} \citep{galclean} algorithm. The clean images can be seen in the middle row. The next issue we address is the orientation of the galaxies. As it has been shown in previous works \citep{Astrovader} this is one of the dominant features to a network and so we rotate all of our galaxies to prevent this from becoming an issue. The last feature we address is the apparent size of the galaxies. We re-scale all of our images to the average petrosian radius of 10 pixels. This will allow the network to focus on the finer details of the images instead of using information encoding the size of the galaxies, which we are not interested in. As our galaxies are all high redshift, we also crop the images to remove as much background as possible. The resulting images can be seen in the bottom row of Fig.\ref{fig:aug}.

\begin{figure*}[hbt]
\resizebox{\hsize}{!}{\includegraphics[clip=true]{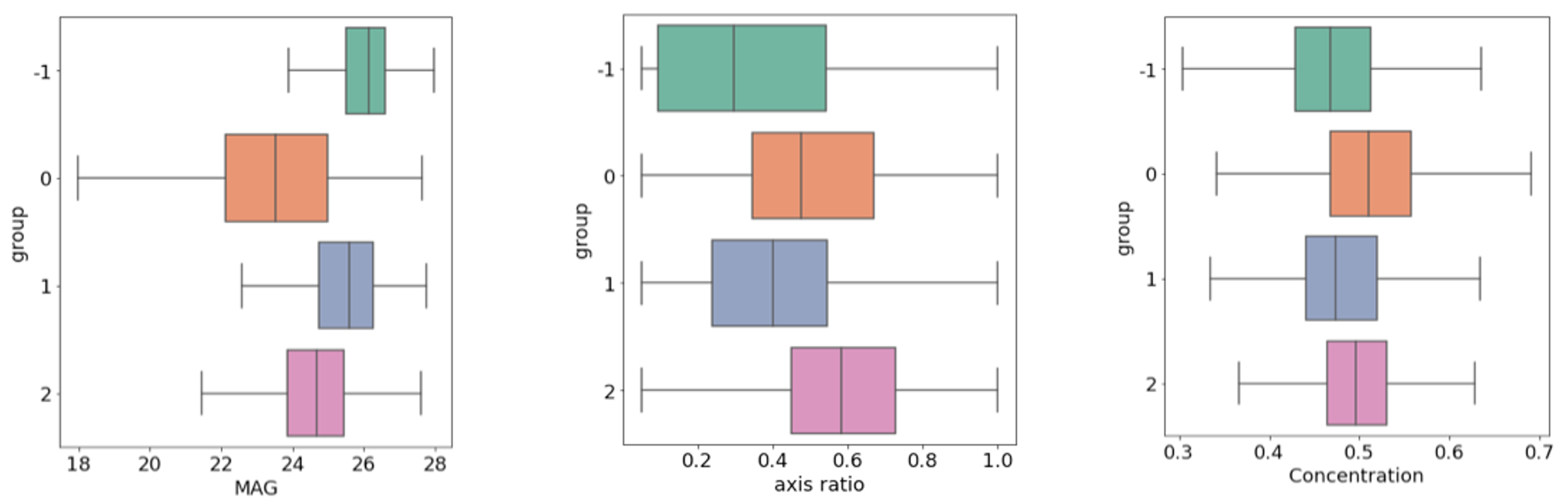}}
\caption{\footnotesize Box plots showing the different physical paramters for the 4 groups found using the density based clustering technique HDBSCAN. Group -1 indicates noise, points that were not close to any dense regions. In total, 9\% of our sample is considered noise.}
\label{fig:groups}
\end{figure*}

\section{Preliminary results}
Utilising our augmented images, we first train the network to encode the images into a small number of features to ensure that our augmentation process has removed the orientation effects from dominating the feature space. The results from this can be seen in Fig. \ref{fig:recon}. It can be seen that the network is able to encode the general shape, elipticity and concentration of our images. This is a good sign as we do not see any orientation effects taking up any of the encoded information. We can also see the effect of information loss, our reconstructions are very smooth and have removed the noise from the images as well.

\subsection{Clustering}
The aim of our work is to be able to separate our galaxies into different clusters based on their intrinsic morphological features that are extracted by our network. We apply two different clustering algorithms to the feature space, a Gaussian based method (Bayesian Gaussian mixture model) and a hierarchical density based algorithm (HDBSCAN). A feature of the type of VAE we are using in this work means that the feature space can deviate from a Gaussian distribution, so we found that using the density based clustering we were able to obtain a better separation of the groups than with the Gaussian mixture model. This density based method found four main groups for our feature space, the properties of the galaxies in each group are shown in Fig.\ref{fig:groups}. We can see that by encoding the data into just three latent dimensions or features, we can already split our galaxies based on axis ratio, magnitude and a slight variation in concentration.

\section{Conclusions and future work}
The next step is to increase the number of features that the network can use to encode the information, whilst still keeping the number of dimensions small enough to be interpretable and comparable to known physical properties. Once we have determined the different populations of galaxies that exist we can then link this with the redshifts of our sample to try to investigate how these groups evolve into the Hubble Sequence we see today.

\begin{acknowledgements}
I am grateful to the members of the EAS 2022 committee and to the organisers of the symposium for allowing me to both present my work, and publish my results in the Memorie della SAIt. CT acknowledges funding from the Science and Technology Facilities Council (STFC)
\end{acknowledgements}

\bibliographystyle{aa}
\bibliography{ref}{}

\end{document}